\documentclass[aps,prl,preprint,superscriptaddress]{revtex4-2}

\usepackage{amsmath,amsfonts,amssymb}
\usepackage{enumitem}
\setlist[enumerate]{itemsep=0mm}

\usepackage[pdftex]{graphicx}
\usepackage{gensymb}
\usepackage{microtype}
\usepackage{bm}
\usepackage[normalem]{ulem}

\usepackage[svgnames]{xcolor}
\usepackage[
	colorlinks=True,linkcolor=DarkRed,citecolor=ForestGreen,urlcolor=MediumBlue,
	pdfstartview=FitH,bookmarks=False,pdfpagemode=UseNone
]{hyperref}

\definecolor{green}{rgb}{0.0, 0.7, 0.0}

\newcommand{\Ir}{IrTe$_2$~}

\begin{document}

\title{\boldmath Bonding states underpinning structural transitions in \Ir observed with micro-ARPES}

\author{C. W. Nicholson}
\email{cw$\_$nicholson@protonmail.com}
\affiliation{University of Fribourg and Fribourg Centre for Nanomaterials, Chemin du Mus\'ee 3, CH-1700 Fribourg, Switzerland}
\affiliation{Fritz-Haber-Institut der Max-Planck-Gesellscahft, Faradayweg 4-6, D-14195 Berlin, Germany}
\author{M. D. Watson}
\affiliation{Diamond Light Source Ltd, Harwell Science and Innovation Campus, Didcot, Oxfordshire, OX110DE, U.K.}
\author{A. Pulkkinen}
\affiliation{University of Fribourg and Fribourg Centre for Nanomaterials, Chemin du Mus\'ee 3, CH-1700 Fribourg, Switzerland}
\affiliation{New Technologies-Research Center, University of West Bohemia, Plze\v n 301 00, Czech Republic}
\author{M. Rumo}
\affiliation{University of Fribourg and Fribourg Centre for Nanomaterials, Chemin du Mus\'ee 3, CH-1700 Fribourg, Switzerland}
\author{G.~Kremer}
\affiliation{University of Fribourg and Fribourg Centre for Nanomaterials, Chemin du Mus\'ee 3, CH-1700 Fribourg, Switzerland}
\affiliation{Institut Jean Lamour, UMR 7198, CNRS-Universit\'e de Lorraine, Campus ARTEM, 2 all\'ee Andr\'e Guinier, BP 50840, 54011 Nancy, France}
\author{K. Y. Ma}
\affiliation{Max Planck Institute for Chemical Physics of Solids, N\"othnitzer Str. 40, 01187 Dresden, Germany}
\author{F. O. von Rohr}
\affiliation{Department of Quantum Matter Physics, University of Geneva, 24 Quai Ernest-Ansermet, CH-1211 Geneva, Switzerland}
\author{C. Cacho}
\affiliation{Diamond Light Source Ltd, Harwell Science and Innovation Campus, Didcot, Oxfordshire, OX110DE, U.K.}
\author{C. Monney}
\email{claude.monney@unifr.ch}
\affiliation{University of Fribourg and Fribourg Centre for Nanomaterials, Chemin du Mus\'ee 3, CH-1700 Fribourg, Switzerland}

\date{\today}

\begin{abstract}

Competing interactions in low-dimensional materials can produce nearly degenerate electronic and structural phases. We investigate the staircase of structural phase transitions in layered \Ir{} for which a number of potential transition mechanisms have been postulated. The spatial coexistence of multiple phases on the micron scale has prevented a detailed analysis of the electronic structure. By exploiting micro-ARPES obtained with synchrotron radiation we extract the electronic structure of the multiple structural phases in \Ir in order to address the mechanism underlying the phase transitions. We find direct evidence of lowered energy states that appear in the low-temperature phases, states previously predicted by \textit{ab initio} calculations and extended here. Our results validate a proposed scenario of bonding and anti-bonding states as the driver of the phase transitions. 

\end{abstract}

\maketitle

The diversity of electronic and magnetic phenomena occurring in quasi-2D transition metal dichalcogenides (TMDs) can largely be traced back to the presence of a partially filled $d$ valence band \cite{Wilson1969}. The relatively narrow $d$-bands, with widths on the same order as the Coulomb interaction, produce ideal conditions for correlated phenomena such as superconductivity \cite{Revolinsky1963, Morosan2006, Sipos2008}, metal-insulator transitions \cite{Fazekas1979}, and charge density waves beyond the Peierls paradigm \cite{Wilson1979a, Cercellier2007, Rossnagel2011}.

\begin{figure}
\includegraphics[width=\columnwidth]{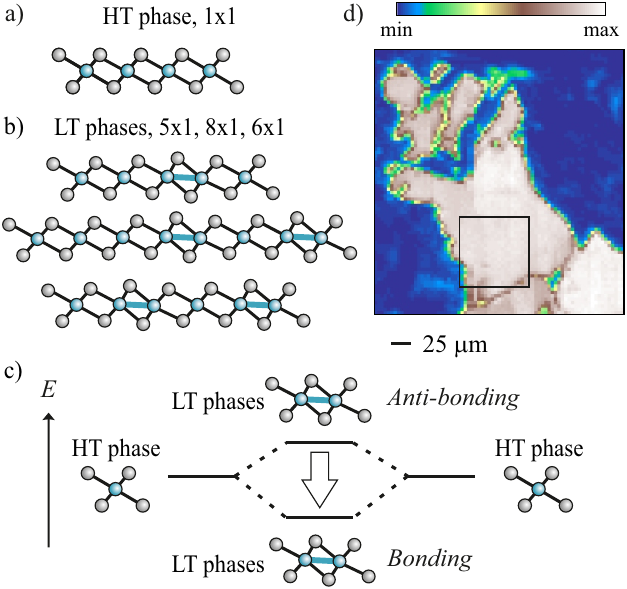}
\caption{a) Side view of the crystal structure of 1$T$-IrTe$_2$ in the high temperature (HT) 1x1 phase, and b) in the various LT phases (top row 5x1, middle row 8x1, bottom row 6x1). c) Schematic bonding energy level diagram in the HT and LT phases, highlighting the formation of bonding states due to the dimer bonds. d) Overview image of the sample obtained with micro-ARPES with a pixel size of 5$\mu$m. Blue color corresponds to the background signal from the epoxy used to glue the sample.}
\label{fig:overview}
\end{figure}

Here we focus on the unusual structural transitions in the well-known TMD IrTe$_2$ involving Ir atoms formally in a 5$d^6$  electronic configuration. Already at 280~K the crystal structure of \Ir transitions from a trigonal 1$T$ phase ($P$-3$m$1) 
to a monoclinic ($P$-1) 5x1x5 superstructure in a first-order fashion \cite{Yang2012}. In the bulk this is followed by a second transition at 180~K into an 8x1x8 phase \cite{Ko2015}. Low-temperature (LT) measurements have revealed a 6x1x6 ground state \cite{Hsu2013, Takubo2018}, which can also be stabilized in Se doped \cite{Oh2013} and uniaxially strained crystals \cite{Nicholson2021}. Although the resistivity is reduced in the LT phases, \Ir remains metallic at all temperatures. Each of the phases below 280~K exhibits a shortening of some of the Ir-Ir distances within the unit cell into effective dimers \cite{Pascut2014a}. The number of these Ir dimers within a unit cell is characteristic to each of the LT phases \cite{Ko2015, Rumo2020} (see Fig.~\ref{fig:overview}a and b). Furthermore, in monolayer samples a novel insulating 2x1 behavior has been observed \cite{Hwang2022a}. Determining the mechanism of the structural transitions is therefore an important task.

Initially proposed weak-coupling scenarios based on Fermi nesting \cite{Yang2012} appear to be incompatible with the dramatic 20 percent reduction in Ir bond lengths later observed in the LT phases \cite{Pascut2014a}. Alternative scenarios include a $J_{\mathrm{eff}}=1/2$ spin-orbit Mott state \cite{Ko2015}, similar to that observed in Sr$_2$IrO$_4$ \cite{Kim2008}, proposed to arise due to a possible Ir 5$d^5$ (Ir$^{4+}$) configuration in the LT phases. A number of works have put forward ideas based on a molecular-type picture involving either depolymerisation between the layers \cite{Oh2013}, or an effective Ir dimerization that produces bonding and anti-bonding 5$d$ states \cite{Pascut2014a, Pascut2014b} (see Fig.~\ref{fig:overview}c) as part of a more complex multi-centre bond \cite{Saleh2020}. Further support for this scenario was obtained by recent high-pressure diffraction measurements, where an observed modification of the Ir-Te angle provides conditions compatible with a ring-shaped bond across multiple sites \cite{Ritschel2021a}. In both the LT and high-pressure calculations, the predicted bonding states appear distinctly split-off to higher binding energies from the density of states (DoS) of the remaining undimerized Ir atoms. Direct experimental confirmation of such states has not been obtained.

Validation or falsification of this scenario therefore requires clear electronic structure data e.g. from angle-resolved photoemission spectroscopy (ARPES). Numerous studies have provided insights into the electronic structure of \Ir \cite{Ootsuki2013, Ootsuki2017, Qian2014, Ko2015, Blake2015, Lee2017a, Rumo2020, Nicholson2021, Bao2021, Rumo2021, Mizokawa2022}. However, extraction of the intrinsic electronic structure is complicated by the spatial coexistence of a ladder of nearly degenerate phases with periodicity $3n+2$ ($n=1, 2, 3,...$) at the surface with 5x1, 8x1 and 6x1 the dominant phases in observations \cite{Hsu2013}. Typical domain sizes ($\sim10~\mu$m \cite{Bao2021}) are beyond the reach of most ARPES setups, with the result that only the 6x1 phase has been unambiguously isolated with strain-ARPES \cite{Nicholson2021}. An outstanding issue is therefore to address the electronic structure in the intrinsic LT structures without extraneous influence. 

\begin{figure*}
\includegraphics[width=\columnwidth]{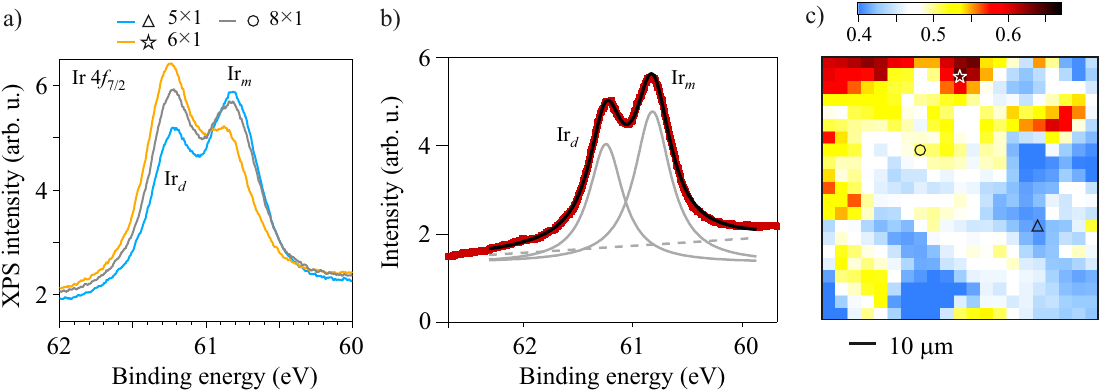}
\caption{a) XPS spectra of the 4$f_{7/2}$ core level obtained from specific domains with a single phase (the individual positions are indicated by markers in (c)).
b) Exemplary fit of the Ir 4$f_{7/2}$ core-level in the 5x1 phase. The fit function comprises of the sum of two Voigt peaks on top of a linear background and a constant offset. Individual component curves and the background are shown. The spectral weight under each peak is extracted and used to plot the ratio of peaks in c).
c) XPS map of the region highlighted (black rectangle) in Fig. \ref{fig:overview}d obtained at 160~K and 130 eV photon energy. The color scale encodes the proportion ($R$) of Ir$_d$ relative to the total population of Ir atoms, $R$ = Ir$_d$/(Ir$_m$+Ir$_d)$. The observed $R$ values for the three spectra shown in (a) are: 0.64 (star), 0.50 (circle) and 0.41 (triangle).}
\label{fig:xps_map}
\end{figure*}

We overcome this challenge by applying micro-ARPES and micro-XPS (X-ray Photoemission Spectroscopy) with synchrotron radiation to \Ir in order to isolate the electronic structure of the surface 5x1, 8x1 and 6x1 phases. Individual phases are identified by their characteristic Ir 4$f$ core level spectra, allowing us to determine the spatial extent of domains on the micrometer scale, and to extract the intrinsic band structure of individual phases with ARPES. We observe split-off states in the DoS of the LT phases, clearly supporting a bonding mechanism for the phase transition.

Single crystals of \Ir were grown using the self-ﬂux method \cite{Jobic1991, Fang2013} and were characterized by magnetic susceptibility and resistivity measurements \cite{Rumo2020}. Micro-ARPES and micro-XPS measurements were carried out on the nano-branch of the I05 beamline at Diamond Light Source with a hemispherical analyzer in a base pressure of low 10$^{-10}$ mbar. For these measurements we selected beam sizes produced by a micro-focussing capillary in order to achieve a spot size of around 6 $\mu$m (FWHM) with high photon flux on the sample. This configuration was chosen to allow efficient probing of the typical domain sizes \cite{Bao2021}. All data were obtained with $p$-polarized light. The ARPES measurements shown here were carried out between 70 -- 90 eV, while XPS was obtained at a photon energy of 130 eV due to the increased Ir 4$f$ cross-section. All crystals were cleaved in vacuum using a ceramic top-post. 

The density functional theory DoS calculations were performed using the Vienna \textit{ab-initio} simulation package (VASP)\cite{Kresse1993, Kresse1994, Kresse1996, Kresse1996a} within the projector augmented wave method \cite{Kresse1999} and the generalized gradient approximation using the Perdew-Burke-Ernzerhof functional \cite{Perdew1996}. The kinetic energy cutoff was set to 400~eV and the Brillouin zone was sampled with the tetrahedron method \cite{Blochl1994} and a $\Gamma$-centered $k$-point grid with spacing $0.08\: \mathrm{\AA^{-1}}$. Spin-orbit interaction was included in all calculations.

A large scale micro-ARPES map of an \Ir sample obtained at 50~K with 5 $\mu$m step size is shown in Fig.~\ref{fig:overview}d. The image is produced by integrating the ARPES intensity over the full detector, and represented according to the color scale shown. We focus on the area highlighted by the black rectangle in order to perform more detailed mapping, as multiple phases are found within the scan area. We initially carried out an XPS map in order to determine the distribution of phases at a sample temperature of 160 K i.e. below the nominal 5x1 - 8x1 transition. Since the ratio of dimer (Ir$_d$) to monomer (Ir$_m$) Iridium atoms is different in each LT phase, each phase has a characteristic ratio $R$ = Ir$_d$/(Ir$_m$+Ir$_d)$ of the integrated intensity under the XPS peak corresponding to those two separate Ir environments \cite{Rumo2020} e.g. in the different Ir 4$f_{7/2}$ peaks. Thus the ratio is $2/5$ (0.4) for the 5x1 phase, $4/8$ (0.5) for the 8x1 phase, and $4/6$ (0.67) in the 6x1 phase (see Fig.~\ref{fig:overview}c). Exemplary individual XPS spectra of the 4$f_{7/2}$ core-level for each LT phases are presented in Fig.~\ref{fig:xps_map}a, revealing the strong increase of the dimer signal at 61.5 eV in the 8x1 and 6x1 phases, in comparison to the 5x1 phase. The ratio is obtained from the XPS data by fitting with a two-peak Voigt function on a linear background and a constant offset (see Fig.~\ref{fig:xps_map}b). The resulting spatial ratio plot of $R$ is shown in Fig.~\ref{fig:xps_map}c. From the color scale it is evident that all three phases coexist at this temperature, which is somewhat surprising given that in the bulk the 5x1 and 8x1 phases do not coexist \cite{Ko2015}. Nevertheless, this is in good agreement with our previous photoemission study indicating coexistence of different phases at low temperature at the surface \cite{Rumo2020}. The XPS spatial map is dominated by regions with $R$ values near 0.4 and 0.5, indicating dominantly 5x1 and 8x1 phases. The gradual variation in the color scale near domain boundaries suggests a partial mixing of phases within the selected spatial resolution. 

\begin{figure*}
\includegraphics[width=\columnwidth]{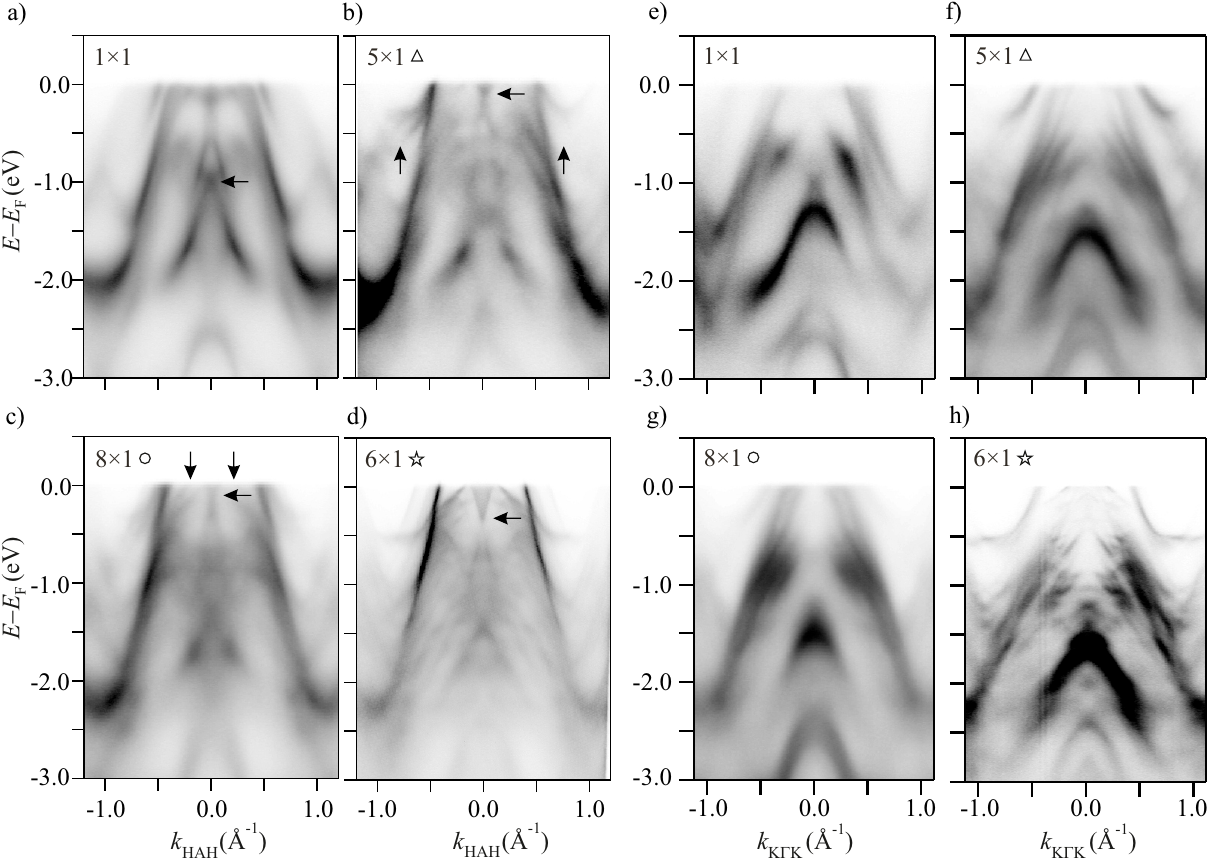}
\caption{Experimental band structure at the bulk A-point (90~eV photon energy) along the high-symmetry direction $HAH$ of the HT Brillouin zone in the a) 1x1, b) 5x1, c) 8x1, and d) 6x1 phase. Arrows mark the key features described in the text. Experimental band structure at the bulk $\Gamma$-point (70~eV photon energy) along the high-symmetry direction $K\Gamma K$ of the HT Brillouin zone in the e) 1x1, f) 5x1, g) 8x1, and h) 6x1 phase. All spectra are obtained at the points marked in Fig.~\ref{fig:xps_map}, and at a temperature of 160~K.}
\label{fig:dispersions}
\end{figure*}

The spatial distribution of phases obtained from the XPS data also allows us to probe the individual LT phases with ARPES; the results obtained at 160~K along the H-A-H direction and the $K-\Gamma$-K direction of the bulk Brillouin zone of the HT phase are shown in Fig.~\ref{fig:dispersions}. A global similarity can be observed between all phases, including the 1x1 phase obtained at room temperature (Fig.~\ref{fig:dispersions}a). Notable features in the 1x1 phase that are carried over to the LT phase include the intense band dispersing from -2~eV at the zone boundary to the Fermi level, and the topological surface state \cite{Bahramy2018b} with an energy of -1~eV at the zone centre, which is shifted to higher binding energies in the LT phases \cite{Rumo2020}. 

However, distinct differences are observed across the energy window in the LT phases, particularly in the region close to the Fermi level. In the 5x1 phase (Fig.~\ref{fig:dispersions}b) mirror symmetry is broken (the corresponding low-symmetry space group is $P$-1), as seen from the states off-centre around -0.3~eV at $k_{HAH} = \pm 0.6$ $\mathrm{\AA}^{-1}$ (see arrows). 
In addition, in each of the LT phases, the bottom of the bulk Dirac states, unoccupied in the 1x1 phase, are now shifted into the occupied states at the zone centre (see the horizontal arrows in panels b,c,d). This feature is similar in the 5x1 and 8x1 (Fig.~\ref{fig:dispersions}b,c) phases, but the hyperbolic dispersion only becomes evident in the 6x1 phase (Fig.~\ref{fig:dispersions}d) as previously observed \cite{Nicholson2021}. A clear difference between the 5x1 and 8x1 phases is observed close to the zone centre. Between the Dirac states and the intense band dispersing to -2~eV, only a single band is observed dispersing to the Fermi level in the 5x1 phase. In contrast the 8x1 phase shows two states dispersing to the Fermi level, which may be a result of the broken inversion symmetry of this phase that is expected to lift the spin degeneracy \cite{Pascut2014a}. For completeness, the 6x1 phase appears identical to that obtained in strained samples \cite{Nicholson2021}. In contrast to a previous study \cite{Ko2015} we do not find evidence for the appearance of a flat band at $\Gamma$ below 280 K and instead find a dispersing band in the expected region along the high symmetry direction $K\Gamma K$ (Fig.~\ref{fig:dispersions}f,g,h). The corresponding ARPES data for the 1x1 phase is shown in Fig.~\ref{fig:dispersions}e. Since a non-dispersing state is expected for the spin-orbit Mott state, we therefore conclude that a Mott phase in \Ir is unlikely. This difference compared to the previous work may be due to the specially chosen spatial resolution used in the current study, removing any influence from a mixture of domains with different orientations or from other phases. 

As mentioned above, the energy of the topological surface state is shifted to higher binding energies compared with the 1x1 phase. The value of the shift is -0.37~eV (5x1), -0.34~eV (8x1), and -0.46~eV (6x1) for the three LT phases. These values are slightly larger, but with the same trend, as previously determined values employing larger beam sizes in ARPES measurements \cite{Rumo2020}. This shift of the surface state implies a lowered energy configuration in the LT states, which foreshadows the observation of bulk bonding states discussed in more detail below. 

\begin{figure}
\includegraphics[width=11cm]{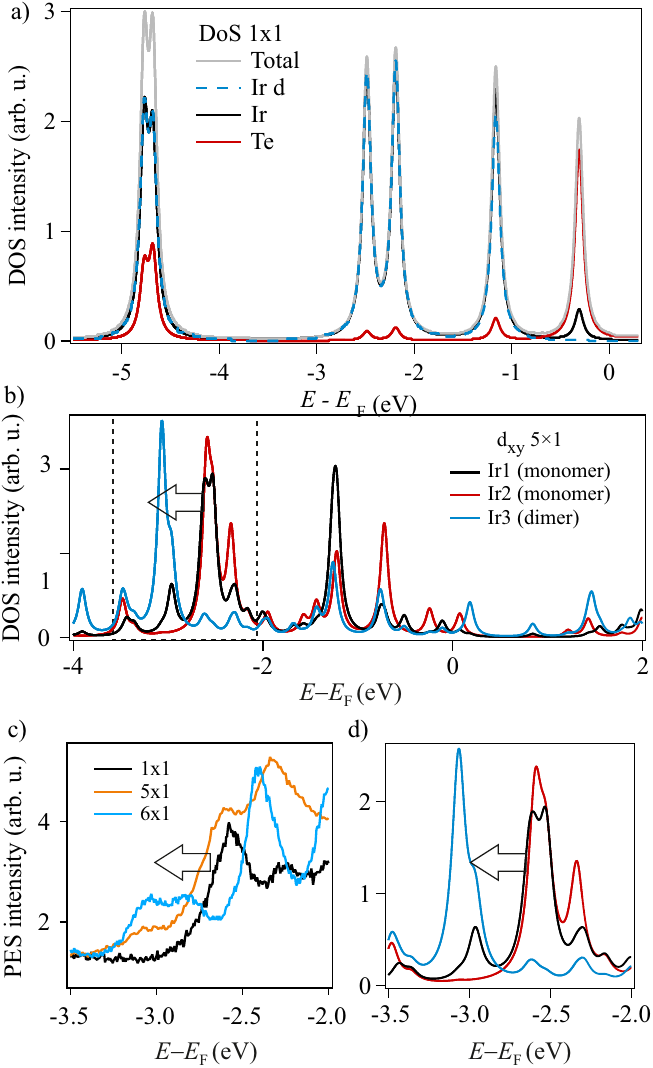}
\caption{a) Calculated partial DoS in the 1x1 phase for the all atoms, for Te atoms only, for Ir atoms only and for the $d$-orbitals of the Ir atoms only. A pseudocubic axis system was used with the $x,y$ axes pointing approximately along the Ir-Te bond directions, such that the Ir $d_{xy}$ orbitals are directed along the dimer bonds. b) Calculated partial DoS in the 5x1 phase for the $d_{xy}$ orbital for the different Ir environments. c) Experimental spectra extracted at $k_{K\Gamma K}=0$ and normalized to the intensity at -3.5~eV in the 1x1, 5x1 and 6x1 phases. Data are obtained at 70~eV, corresponding to the bulk $\Gamma$-point. d) Magnification of the region shown in c) for the calculated partial DoS.}
\label{fig:DOS}
\end{figure}

Given the suggestive behavior of the surface states, we investigated the split-off bonding DoS predicted by bulk DFT calculations in the region of -3~eV binding energy \cite{Pascut2014a}. In order to better compare the momentum-resolved experimental data with the predicted DoS, we have calculated the orbital-resolved DoS within a small momentum range close to the zone centre ($\Gamma$-point) to avoid including the highly dispersive states at higher momenta present at all temperatures. We use a pseudocubic axis system with the $x,y$ axes pointing approximately along the Ir-Te bond directions, such that the Ir $d_{xy}$ orbitals are directed along the dimer bonds. The 1x1 phase is dominated by the Ir $d$-orbitals as can be seen in Fig.~\ref{fig:DOS}a. In the following, we focus on the $d_{xy}$ orbital in the 5x1 phase, since the $d_{xy}$ orbitals overlap directly in the formation of the Ir dimers, resulting in large binding energy shifts, as previously addressed by Pascut $et~al.$ \cite{Pascut2014a}. The 5x1 phase is chosen due its smaller unit cell compared with the 8x1, and the fact that structural data for the intrinsic 6x1 phase is not available \cite{Oh2013, Pascut2014a, Ritschel2021a}. The three curves in Fig.~\ref{fig:DOS}b reflect the three inequivalent Ir sites in the 5x1 phase: dimerized atoms, and two species of monomers depending on their proximity to the dimers. A distinct shift of the dimer DoS to 0.5~eV lower binding energy compared with the monomers is observed. We note here that a similar behavior is observed in the $d_{yz}$ and $d_{xz}$ orbitals. These split-off states are the fingerprints of bonding state formation as proposed by previous works \cite{Pascut2014a,Saleh2020}. Corresponding anti-bonding states are predicted above the Fermi level at around 1.5~eV, see Ref. \cite{Saleh2020} and Fig.~\ref{fig:DOS}b. Our result is similar to the previous calculations, although it displays a clearer effect due to the selective momentum range applied. We note that similar split-off states are expected also at other binding energies below the range currently under investigation \cite{Saleh2020}.

To compare directly with the calculated DoS, we extract an energy slice at the zone centre ($\Gamma$-point) from ARPES measurements, presented in Fig.~\ref{fig:DOS}c. The energy range is chosen to encompass the DoS states highlighted in Fig.~\ref{fig:DOS}b which have already been shown to be of most relevance. In the 1x1 data (black curve) the tail of the DoS extends to $-3$~eV, but no peaks are observed in this region. In contrast, in the 5x1 phase a clear peak is evident at $-3.05$~eV, with a difference in energy of $-0.46$~eV compared with the lowest energy peak in the 1x1 phase. This peak becomes even more pronounced in the 6x1 phase, with a double peak structure forming that mirrors that of the DoS calculation. The ARPES data therefore provides direct evidence for states split-off from the 1x1 DoS associated with the LT phases. 

The observation of split-off states in the LT phases, at energies well separated from the 1x1 DoS, is in excellent agreement with the prediction from our momentum-selective calculations, and fully compatible with the previous momentum-integrated calculations. The close correspondence of experiment and theory therefore provides direct evidence for bonding state formation tied to the structural transition, validating a scenario of strongly bonded atoms as the driver of the structural phase transition \cite{Pascut2014a, Saleh2020}. Indeed, such a scenario provides a natural alternative to the formation of an electronic gap, which is not observed in \Ir and where a metallic, albeit reduced, DoS at the Fermi level is always observed in the LT phases. The large energetic shifts of 0.5~eV in \Ir are comparable but larger than the typical electronic energy gains in quasi-2D CDW materials \cite{Rossnagel2011}.

The observed split-off states are clearly associated with bulk Ir 5$d$ states, and are far-removed from the surface states in energy. This confirms that the electronic structure differences observed here with ARPES are representative of the bulk transition. We note that this situation may not hold in very thin samples, where a number of additional phenomena are observed, including an enhanced transition temperature for the 5x1 phase \cite{Song2021c} and a coexistence of structural 5x1 order and superconductivity \cite{Park2021a}. These observations further imply that the out-of-plane bonding is involved in the phase transition. Indeed, the role of the Te-Te bond was already discussed in early works to explain the anomalously small $c/a$ ratio \cite{Jobic1991, Jobic1992}, and the results of Se-doping \cite{Oh2013}. In fact, a reduction of interlayer bond strength is an additional consequence of the multi-centre bond formed in the LT phases \cite{Saleh2020}, and is supported by the reduced interlayer coupling observed in ARPES of the 6x1 phase \cite{Nicholson2021}. This highlights that the picture of the dimer motif forming bonding states provides a coherent description of the observed data for the structural transitions in \Ir.

In summary, we used synchrotron micro-ARPES to separate the intrinsic electronic structure of the micro-scale LT phases in IrTe$_2$. The rich structures unveiled display mirror symmetry breaking in the 5x1 phase, and possible features of inversion symmetry breaking in the 8x1 phase. The structure of the 6x1 phase appears to be equivalent to that stabilized under uniaxial strain. Disentangling the various phases allowed us to resolve split-off bonding states that appear in the LT phases at around -3~eV binding energy, as previously predicted based on structural models. Our DoS calculations confirms the appearance of these states in terms of $d_{xy}$, $d_{yz}$, $d_{xz}$ and $d_{x^{2}-y^{2}}$ orbitals. Our results provide experimental evidence for a mechanism based on short dimer bonds as the energetic driver of the structural transitions in layered IrTe$_2$.

\section{Acknowledgments}
This project was supported by the Swiss National Science Foundation (SNSF) Grant No. $P00P2\_170597$. 
A.P. would like to thank the QM4ST project with Reg. No. CZ.02.01.01$/$00$/$22\_008$/$0004572, cofunded by the ERDF as part of the M\v SMT.
We thank Diamond Light Source for acces to beamline I05 under proposal no. SI25906.

\end{document}